\documentclass[a4paper,11pt]{JHEP3}
\usepackage{amsmath}
\usepackage{epsf, cite, amssymb}
\usepackage{epsfig}
\title{Comments on I1-branes}
\author{Ling-Yan Hung \\ Department of Applied Mathematics and
  Theoretical Physics, \\ Wilberforce Road, Cambridge CB3 0WA, UK
\\ E-mail: \email{lyh20@damtp.cam.ac.uk} }

\abstract{We explore the supergravity solution of D5-branes
  intersecting as an I1-brane. In a suitable near-horizon limit the
  geometry is in qualitative agreement with that found in the
  microscopic open-string analysis as well as the NS5-brane analysis
  of Itzhaki, Kutasov and Seiberg. In particular, the $ISO(1,1)$ Lorentz
  symmetry of the intersection domain is enhanced to $ISO(1,2)$. The
  discussion is generalised to the T-dual configuration of a D4-brane
  intersecting a D6-brane. In this case the $ISO(1,1)$ symmetry is not enhanced.
This is true both in the supergravity approximation to the weakly coupled
  string theory and to the M-theory limit.}

 \preprint{DAMTP-2006-121}

\begin{document}

\section{Introduction}
Gauge anomalies play a key role in the structure of quantum field
theory and string theory. They also enter into
curious phenomena in some brane configurations. It was recently
observed in \cite{IKS} and independently in \cite{Lin} that close to
the $1+1$ intersection region of intersecting five 
branes (an I1-brane) Poincare symmetry is enhanced from $ISO(1,1)$ to
$ISO(1,2)$. This was deduced in two ways. The first was a microscopic
weakly-coupled open string analysis, involving anomaly inflow from the
branes to the intersecting region. The second involved the
supergravity limit of the S-dual brane configuration (i.e. NS5-branes)
in the near horizon limit. In fact the algebra of the enhanced
supersymmetries is presented in \cite{Lin}.
In this note we analyse the supergravity limit of the branes
directly in their D5-brane description. 
In this case special care is needed in taking the near horizon
limit. Although the
supergravity solution is applicable only in the large $N$ limit, where the
number of branes in both stacks is large, and in
a restricted range of the radial coordinates, the general picture
that symmetry is enhanced is unchanged. 

Then we move on to studying the T-dual configuration, namely a stack of
D6's intersecting another stack of D4's in $1+1$ dimensions. 
Following the analysis in \cite{IKS} we find that in this case
there is no symmetry enhancement in both the gauge theory limit and
the supergravity limit. In the gauge theory limit the anomaly
inflow mechanism is still at work so that the chiral fermions are
still displaced from the intersection region. However they are pushed into
the bulk of D6 and D4, respectively, with different radial
dependences. This spoils the level-rank duality\cite{level_rank}
argument so that
Poincare symmetry enhancement is absent. This is confirmed in the
supergravity limit where we find that angle deficits develop in the
near horizon limit which prevents symmetry enhancement. We will note
that this is also true of the M-theory description.

The organisation of this paper is as follows:
Section 2 gives an overview of the
results of \cite{IKS}. 
Section 3 discusses the supergravity
interpretation of the intersecting D5's. The  generalisation to the
D6-D4 system is described in section 4. 
Further comments are made in the last section, including a discussion
of the M-theory limit of the D4/D6 system.

\section{Overview of I-brane dynamics}\label{Ibraneinflow}

Consider a BPS configuration of orthogonally intersecting branes,
such that the number of relatively transverse dimensions is a
multiple of four and that the intersection domain has 
$4k+2$ dimensions. The massless spectrum consists chiral fermions that
arise from open strings connecting a brane from each stack of
intersecting branes. They give rise to gauge and gravitational
anomalies from the perspective of the field theory in the intersection
domain. This anomaly is cancelled by anomaly inflow from the rest of the brane\cite{green}, regarded here as the bulk. This mechanism implies that the D-brane
world-volume action contains Chern-Simons terms of the form

\begin{equation}\label{coupling}
\mu\int_{M_{Dp}} C\wedge ch_N(F)\sqrt{\hat{A}(R)},
\end{equation} 
where $C$ refers to the RR-forms, $ch_{r}(F) = Tr_{r}(exp(\frac{iF}{2\pi}))$ and
$\hat{A}(R)$ is the A-roof genus. The integral picks out the forms
such that the wedge product is a $p+1$ form on the D$p$ brane world-volume.
This mechanism played a crucial role in the microscopic analysis in
\cite{IKS}, which we will now briefly review.

\subsection{Open string perspective of intersecting D5 branes}
Consider $N_1$ D5 branes intersecting another set of $N_2$ D5's
orthogonally over 1+1 dimensions. Suppose the first set lies along $x^\mu, \mu \in \{0,1,2,3,4,5\}$ and the second set along $\mu \in \{0,1,6,7,8,9\}$. There are altogether 8 relatively transverse directions. The system preserves a quarter supersymmetries i.e. eight supercharges. 
\begin{equation}
\Gamma^{012345} \epsilon_{R} = \epsilon_{L}, \qquad \Gamma^{016789} \epsilon_{R} = \epsilon_{L}.
\end{equation}
In the low energy limit the $5-5'$ open strings essentially live in the intersection region and the field theory of interest is 1+1 dimensional. From the perspective of the 1+1 dimensional theory, the supercharges preserved are chiral satisfying
\begin{equation}
\Gamma^{01}\epsilon_{L,R} = \epsilon_{L,R}.
\end{equation}
The $5-5'$ sector also contains massless chiral fermions. They
originate from the RR zero modes along $x^0$ and $x^1$. After GSO
projection we are left with one chiral fermion in the representation
$(N_1,\bar{N_2})$ of the gauge group $U(N_1)\times U(N_2)$ and so the field theory in the intersection is anomalous.  However the theory including the intersection and bulk is
anomaly free as discussed earlier.
The total tree-level Lagrangian for the whole system is a sum of the kinetic
terms of the gauge fields 
and the chiral fermion and the Chern-Simons terms. 
The fermions are then
integrated out to give a non-local action. 
The low energy limit is taken such that only the S-waves in the two
3-spheres in the bulk of each of the two sets of branes are included. 

The resulting action is quadratic in the fields and is explicitly
anomaly free. Equations of motion obtained from this
effective action describe wave functions for the gauge fields that are
displaced away from the intersection region. For the simple case where $N_1 =
N_2 = 1$, the solution is \cite{IKS}

\begin{equation}
F^{(i)}_{u_i\pm} = \frac{h_\pm}{u_i^3} e^{\pm \frac{g_i^2}{u_i^2}},
\end{equation}
where $i \in \{1,2\}$. $F^{(i)} =dA^{(i)}$ are the gauge field
strengths of the respective branes with effective gauge coupling $g_i
\sim g_sl_s^2$, and
$u_i$ are the two radial directions away from the intersection region
into the bulk of the branes. 
This implies that the chiral fermion is also displaced away from the
intersection region. This general picture is not altered even if more
coincident branes are considered such that the gauge theory in the
intersection becomes a non-abelian theory. In general, where there
are $N_1$ and $N_2$ branes in each stack respectively, 
we are left with a $SU(N_1)_{N_2}$ Chern-Simons theory at $u_1
\sim g_1\sqrt{N_2}$ and another $SU(N_2)_{N_1}$ Chern-Simons theory at 
$u_2 \sim g_2\sqrt{N_1}$ which, by level-rank duality\cite{level_rank}, are the
same. This level-rank duality is related to the modular invariance of
the torus of the gravity dual \cite{Lin}. Close
to the intersection region the two radial
directions cannot be distinguished. 
Poincare symmetry
is enhanced from $ISO(1,1)$ to $ISO(1,2)$ and the number of conserved
supercharges is also enhanced from 8 to 16.   

\subsection{Closed string perspective}
The same system can also be analysed by considering it as a
black-brane solution to type IIB supergravity in the near horizon
limit. It happens that for intersecting branes with exactly eight
relatively transverse dimensions,  a fully localised solution is
known. Moreover it is actually more convenient to analyse the S-dual
configuration. This is because the supergravity solution for the
D-branes has a non-constant dilaton which grows indefinitely as we
move away from the branes. As a result it ceases to be a good
approximation away from the intersection. More importantly, in the
D-brane description the 
number of branes $N_1$ and 
$N_2$ has to be large in order that the solution has a small
enough curvature to be valid as a supergravity approximation. These constrain
severely the regime where the solution is valid. We will
return to D5-brane description in section 3. On the other hand
switching to the S-dual picture with intersecting NS5-branes, 
the solution turns out to be an exact background of a conformal
world-sheet theory\cite{CHS} and so the geometry obtained is not
restricted to the supergravity approximation. The solution is valid
for all $N_1, N_2>1$. Following the notation in \cite{IKS}, the metric
of the configuration is given by 
\begin{equation}
ds^2 = -(dx^0)^2 + (dx^1)^2 + f_1(v)(dv^2 + v^2d\Omega^2_v) +  f_2(u)(du^2 + u^2d\Omega^2_u),
\end{equation}
where
\begin{eqnarray}
y      & = &(x^2,x^3,x^4,x^5),  \nonumber \\
z      & = &(x^6,x^7,x^8,x^9), \nonumber  \\
\Phi  & = & \Phi_1(y_2) + \Phi_2(y_1) ,
\end{eqnarray}
and
\begin{eqnarray}
e^{2(\Phi_1 - \Phi_1(\infty))} = f_1(v=|z|) &=&1 + \sum_{n=1}^{N_1} \frac{l_s^2}{|z-z_n|^2},  \nonumber \\
e^{2(\Phi_2 - \Phi_2(\infty))} =f_2(u=|y|) & = &1 + \sum_{n=1}^{N_2} \frac{l_s^2}{|y-y_n|^2}.
\end{eqnarray}
The $y_{n}$'s and $z_n$'s give the position of the branes. In the case
under consideration $y_n=z_n =
0$. The near horizon limit is obtained by keeping
$v/\exp(\phi_1(\infty))$  and $u/\exp(\phi_2(\infty))$ fixed while
letting $\exp(\phi_1(\infty)), \exp(\phi_2(\infty)) \to 0$. This
essentially means that  we can drop the constant in front of the term
$l_s^2/|z-z_n|^2$,  
\begin{equation}
e^{2(\Phi_1 - \Phi_1(\infty))} = f_1(v=|z|) \sim  \frac{N_1l_s^2}{|z-z_n|^2},
\end{equation}
and similarly for $\exp(2(\Phi_2 - \Phi_2(\infty)))$.
Making a change of coordinates
\begin{equation}\label{coordchange}
\phi_1 = \sqrt{2k_1}\ln v, \qquad \phi_2 = \sqrt{2k_2}\ln u, 
\end{equation}
where $2k_i=N_il_s^2 $, the resulting metric is
\begin{equation}
ds^2 =  -(dx^0)^2 + (dx^1)^2 + d\phi_1^2 +  d\phi_2^2 + 2k_2d \Omega^2_u +2k_1d\Omega^2_v . 
\end{equation}
The metric looks flat in $x^0,x^1,\phi_1,\phi_2$. However the
dilaton depends on $\phi_1,\phi_2$ and so the background is not
invariant under general $ISO(1,3)$ rotations. Consider a further
coordinate change given by \cite{IKS, Lin}
\begin{equation}
Q\phi  = Q_1\phi_1+Q_2\phi_2, \qquad Qx^2  = Q_2\phi_1-Q_1\phi_2,
\end{equation}
where 
\begin{equation}
Q_i    = \sqrt{\frac{2}{k_i}}, \qquad Q = \sqrt{\frac{2}{k}}, \qquad 
\frac{1}{k} = \frac{1}{k_1} + \frac{1}{k_2}.
\end{equation}
Then we end up with the same metric but now the dilaton depends only
on $\phi$ and so the background is invariant under general $ISO(1,2)$
rotations. The near horizon geometry has an enhanced Poincare symmetry as if an extra dimension has grown out from the intersection region as in the weak coupling description.

\section{The D5 supergravity description}
While the supergravity solution for the intersecting D5-brane
configuration is only valid in a particular region in coordinate space
and within a smaller regime of the couplings, it is nevertheless
interesting to take a closer look at it. The fully localised
solution for D5's intersecting over 1+1 dimensions
is given (in string frame) by 
\begin{equation}
ds^2 = (H_1H_2)^{-\frac{1}{2}}(-(dx^0)^2 + (dx^1)^2  + H_1 (du^2 + u^2d\Omega_u^2)+ H_2(dv^2 + v^2d\Omega_v^2)), \nonumber
\end{equation}
\begin{eqnarray}
e^{(\Phi - \Phi(\infty))} &=& e^{\Phi_1}e^{\Phi_2}, \nonumber \\
e^{\Phi_i} &=& H_i^{-\frac{1}{2}}, \nonumber\\
H_1 &=& 1 + \frac{d_5g^2_{5}N_1}{u^2}, \nonumber \\ 
H_2 &=& 1 + \frac{d_5g^2_{5}N_2}{v^2}, \nonumber \\
g_{5}^2 &=& (2\pi)^3g_s\alpha',
\end{eqnarray}
where $d_5$ is some constant whose value is defined in \cite{Mal}. As
in AdS/CFT we need to take the low energy limit which is the near horizon limit from the supergravity perspective. The Maldacena limit is however not the appropriate limit to be taken here. In the Maldacena limit where we let $\alpha' \to 0$ while keeping $g_5^2$ and $u/\alpha'$ and $v/\alpha'$ fixed \cite{Mal}, both the string coupling and $ds^2$ in units of string length tend to zero. In order to keep these two quantities finite while taking the near horizon limit, we keep instead $U = u/\sqrt{\alpha'}$ and $V= v/\sqrt{\alpha'}$ fixed. This limit is in fact the S-dual of that taken for the intersecting NS5 branes as described in the previous section. This can be seen by noting that under S-duality
\begin{eqnarray}
\tilde{g_s} & = & \frac{1}{g_s}    \nonumber \\
\tilde{\alpha'}  & = & g_s\alpha'  \nonumber\\
\end{eqnarray}

In this limit the metric becomes
\begin{equation}
e^\Phi = \frac{UV}{d_5\sqrt{N_1N_2}}, \nonumber
\end{equation}
\begin{equation}
ds^2 = \alpha' (\frac{UV}{d_5g^2_5\sqrt{N_1N_2}})(-(dx^0)^2 + (dx^1)^2  + d\phi_1^2 + d\phi_2^2 + d_5g^2_5(N_1d\Omega_u^2+ N_2d\Omega_v^2)),
\end{equation}
where 
\begin{equation}
\phi_1 = \sqrt{g^2_5N_1d_5}\ln U, \qquad \phi_2 = \sqrt{g^2_5N_2d_5}\ln V.
\end{equation}
Now we perform the same change of coordinates as in (\ref{coordchange})
\begin{eqnarray}
\sqrt{\frac{1}{N}}\Omega = \frac{\phi_1}{\sqrt{N_1}} +
\frac{\phi_2}{\sqrt{N_2}}, \qquad \sqrt{\frac{1}{N}}x^2 =
\frac{\phi_1}{\sqrt{N_2}} - 
\frac{\phi_2}{\sqrt{N_1}}
\end{eqnarray} 
where
\begin{equation}
\frac{1}{N} = \frac{1}{N_1} + \frac{1}{N_2}.
\end{equation}
The solution reduces to
\begin{equation}
ds^2 = \alpha' (\frac{e^{\Omega/\sqrt{g^2_5d_5N}}}{d_5g^2_5\sqrt{N_1N_2}})(-(dx^0)^2 + (dx^1)^2  + (dx^2)^2 + d\Omega^2 + d_5g^2_5(N_1d\Omega_u^2+ N_2d\Omega_v^2))
\end{equation}
\begin{eqnarray}
e^\Phi  &=& \frac{e^{\Omega/\sqrt{g^2_5d_5N}}}{d_5\sqrt{N_1N_2}}
\end{eqnarray}
So we see that the solution exhibits enhanced Poincare symmetry, just
as it did in the regimes considered in \cite{IKS}

The solution is a supergravity approximation. Therefore it is only valid in the region where both the string coupling and the curvature are small.
The string coupling is small when  $UV \ll \sqrt{N_1N_2}$. Beyond that
we have to go over to the S-dual picture. The Ricci scalar of the
metric is approximately $\sim 1/UV$ and this is small for $UV\gg 1$.  Beyond that we need to go over to the gauge theory picture.

\section{The D6-D4 system}
It is interesting to T-dualise the intersecting D5-branes along one of
the relatively transverse directions to obtain the D6-D4 system. The
branes again intersect over $1+1$ dimensions and in the same way there
is a chiral fermion in the intersection region, whose chiral anomalies
have to be cancelled by the anomaly inflow mechanism via coupling with
the bulk fields through the Chern-Simons terms. However, the symmetry
enhancement observed in the intersection region in the system of
5-branes does not occur in the T-dual picture.  
This can be seen both from weak coupling gauge theory analysis and
from the supergravity limit.  
\subsection{Supergravity solution of D6-D4 system}
System of intersecting branes having eight relatively transverse directions without any totally
transverse directions, is a special case where
a completely localised supergravity solution can be
found\cite{Edelstein, Smith}. Suppose the D4 is 
aligned along $x^{1,2,3,4}$ and D6 along $x^{1,5,6,7,8,9}$. The
supergravity solution of the system is given by
\begin{equation}
ds^2 = H^{-\frac{1}{2}}_6H^{-\frac{1}{2}}_4 \{-(dx^0)^2 + (dx^1)^2 + H_4(du^2+ u^2d\Omega_5^2) + H_6(dv^2+ v^2d\Omega_2^2) \} .
\end{equation}
where
\begin{eqnarray}
z &=& \{x^{2,3,4}\}, \qquad y = \{x^{5,6,7,8,9}\}, \nonumber \\
u &=& |y|, \qquad \qquad v = |z|, \nonumber\\
H_4 &=& 1+\frac{d_4g_sl_s^3N_4}{u^3}, \qquad H_6 =
1+\frac{d_6g_sl_sN_6}{v}, \qquad e^{(\Phi-\Phi(\infty))} =
H_6^{-\frac{3}{4}}H_4^{-\frac{1}{4}}, 
\end{eqnarray}
This solution is valid in the small dilaton limit. As we move away
from the branes the dilaton grows and the 11th dimension becomes
important. The M-theory picture would then be the appropriate
description in which the 
D6-brane becomes a Kaluza-Klein monopole and the D4-brane becomes an M5
brane wrapping on a circle. Consider first the perturbative closed
string limit where the dilaton is small. We take the near horizon
limit as in the case for the D5-branes by sending $\alpha'$ to zero
while keeping the ratio $U= r/\sqrt{\alpha'}$ and $g_s\alpha'$ fixed.  In this
limit we can drop the constant, 1, in $H_4$ and $H_6$. Defining $k_1
= d_6g_sl_sN_6$ and $k_2 = d_4g_sl_s^3N_4$, the metric becomes
\begin{equation}
ds^2 = H^{-\frac{1}{2}}_6H^{-\frac{1}{2}}_4 \{-(dx^0)^2 + (dx^1)^2 +
\frac{k_2}{u^3}du^2+ \frac{k_2}{u}d\Omega_5^2 + \frac{k_1}{v}dv^2+
k_1vd\Omega_2^2) \}. 
\end{equation}
Under a change of coordinates
\begin{equation}
\phi_1 = 2\sqrt{k_1v}, \qquad \phi_2 = -2\sqrt{\frac{k_2}{u}},
\end{equation}
the metric can be written as
\begin{equation}
ds^2 = H^{-\frac{1}{2}}_6H^{-\frac{1}{2}}_4 \{-(dx^0)^2 + (dx^1)^2 + d\phi_2^2+ \frac{\phi_2^2}{4}d\Omega_5^2 + d\phi_1^2+ \frac{\phi_1^2}{4}d\Omega_2^2  \}.
\end{equation}
The part with $d\phi_2^2+ \frac{\phi_2^2}{4}d\Omega_5^2 $ is almost a
flat metric except the factor of 4 in the second term, which cannot be
scaled away by rescaling $\phi_2$. This gives a space with an angle
deficit. This applies also to $d\phi_1^2+
\frac{\phi_1^2}{4}d\Omega_2^2$. Therefore the space does not exhibit
enhanced Poincare symmetry in the near horizon limit.  

\subsection{Gauge field theory analysis}
A similar analysis as for the intersecting D5 case is carried out
here. The calculation is essentially the same. Basically we need to
solve the equations of motion of the effective Lagrangian after
integrating out the chiral fermion that resides in the intersection
region. The effective Lagrangian is  similar to that of the
intersecting D5 case, except for some slight modifications. Taking
only one D6 and one D4, applying again the S-wave approximation where
we ignore all angular dependence in the brane bulk, we have 
\begin{eqnarray}
g^2_{\textrm{Dp}} &=& \frac{g_s^2}{(2\pi\alpha')\mu_p}, \nonumber \\
\mu_p^2  &=& \frac{\pi}{\kappa_{10}^2} (4\pi^2\alpha')^{3-p},  \nonumber \\
L_{\textrm{total}} &=& L_{\textrm{kin}} + L_{\textrm{ferm}} + L_{\textrm{cs}} + L_{\textrm{mix}},\nonumber \\
L_{\textrm{d6kin}}      &=& \frac{V_{S^5}}{g^2_{\textrm{D6}}}\int du u^4 [\frac{1}{2}(F^{(6)}_{+-})^2 - (F^{(6)}_{+u}F^{(6)}_{-u})], \nonumber  \\
L_{\textrm{d4kin}}      &=& \frac{V_{S^2}}{g^2_{\textrm{D4}}}\int dv v^2 [\frac{1}{2}(F^{(4)}_{+-})^2 - (F^{(4)}_{+v}F^{(4)}_{-v})], \nonumber
\end{eqnarray}
\begin{equation}
L_{\textrm{ferm}} = (A^{(6)}_+(0)- A^{(4)}_+(0))\frac{\partial_-}{\partial_+}(A^{(6)}_+(0)- A^{(4)}_+(0)) - (A^{(6)}_+(0)- A^{(4)}_+(0))(A^{(6)}_-(0)- A^{(4)}_-(0)).
\end{equation}
The relevant Chern-Simons coupling to the bulk fields are
\begin{equation}
\int_{M_{6+1}}\frac{1}{\mu_6} H_4 \wedge A^{(6)}\wedge F^{(6)} + H_6 \wedge A^{(6)} + \int_{M_{4+1}} \frac{1}{\mu_4} H_2\wedge A^{(4)} \wedge F^{(4)} + H_4 \wedge A^{(4)}.
 \end{equation}
The equations of motion of the RR forms are
\begin{eqnarray}
dH_2 &=& - \delta(789), \qquad dH_4 = -\delta(23456) -F^{(6)}\wedge
\delta(789) \nonumber \\ 
dH_6 &=& F^{(4)}\wedge\delta(23456) + \frac{1}{\mu_6}F^{(6)}\wedge F^{(6)}\wedge \delta (789).
\end{eqnarray}
Taking into account
\begin{equation}
\int_{S^{p+2}} H_{p+2} = \mu_{6-p}N_{6-p},
\end{equation}
for $p\in \{4,6\}$,
and substituting in the equations of motion of the RR forms, we obtain 
\begin{eqnarray}
L_{\textrm{cs}} &=& -\int du A^{(6)}_+F^{(6)}_{-u} + A^{(6)}_-F^{(6)}_{u+} + A^{(6)}_uF^{(6)}_{+-}  + dv (u \to v, 6\to 4), \nonumber \\
L_{\textrm{mix}} &=& - \left(F^{(6)}_{+-}\int dv A^{(4)}_{v} + F^{(4)}_{+-}\int du A^{(6)}_{u} \right).
\end{eqnarray}

The equations of motion obtained from the Lagrangian by varying the gauge field $A^{(6)}$ at arbitrary $u$ are given by
\begin{eqnarray}
\frac{V_{S^5}}{g_6^2}[\partial_{u}(u^4F^{(6)}_{u-}) + \partial_{-}F_{+-}u^4] -2F_{u-} &=& 0, \nonumber \\
\frac{V_{S^5}}{g_6^2}\partial_u(u^4F^{(6)}) +2F^{(6)}_{u+} + \frac{V_{S^5}u^4}{g_6^2}\partial_+F^{(6)}_{-+} &=& 0, \nonumber \\
2F^{(6)} - F^{(4)}_{+-} - \frac{V_{S^5}u^4}{g_6^2} (\partial_-F^{(6)}_{u+} + \partial_+F^{(6)}_{u-}) &=& 0.
\end{eqnarray}

Similar equations can be obtained by varying $A^{(4)}$, except that we
change all the radial dependence from $u^4$ to $v^2$ and all radial
partial derivatives are taken with respect to $v$. As in eq.(2.18) in
\cite{IKS} the effect of the Chern-Simons terms make the $F_{+-}$ mode
massive. Since we are interested only in the massless modes we can set them to
zero and we are left with
\begin{equation}
F^{(6)}_{u\pm} =
\frac{h^{(6)_\pm}}{u^4}e^{(\pm\frac{2g_6^2}{3V_{S^5}u^3})}, \qquad
F^{(4)}_{v\pm} = \frac{h^{(4)_\pm}}{v^2}e^{(\pm\frac{2g_4^2}{V_{S^2}v})}.
\end{equation}
The calculation shows that the equations of motion
cease to be symmetric between $A^{(6)}$ and $A^{(4)}$ and the gauge
fields are
displaced away from the intersection region with different dependences
on the radial coordinates $u$ and $v$. As a result the theory distinguishes the two radial directions and so we do not get two
identical theories for $u>0$ and $v>0$. Therefore enhanced Poincare
symmetry should be absent, confirming the analysis in the supergravity limit.

\section{Conclusions}
We have seen that the curious enhancement of Poincare
symmetry observed in \cite{IKS,Lin} close to the intersection region of 
intersecting D5-branes is reproduced in the supergravity limit, although the
supergravity solution is only valid over a limited range of the radial
variables and in the large $N$ limit. Special
attention is also needed in taking the double scaling limit in order to
scale to the intersection domain. 
However, this symmetry enhancement is absent in
the T-dual picture in which a D6-brane intersects a D4-brane. From the
open string analysis, this 
manifests itself as different dependences on the radial variables in
the displacement of the chiral fermions from the intersection. From
the supergravity solution in the near horizon limit angle deficits
develop in the brane bulks, which prevent any symmetry enhancement. We
have only treated the supergravity limit where $g_s$ is small. To
consider the strongly coupled picture, the eleventh
dimension in M-theory would emerge and we would need an M-theory
description of the system in which D6-branes are Kaluza-Klein monopoles in
M-theory and the D4-branes are M5 branes compactified on a circle. In
the near horizon limit of $N$ KK monopoles, the geometry reduces
to an ALE space and the transverse four dimensional space can be
described by an orbifold $C^2/\bf{Z_N}$\cite{Smith}. M5 branes can be
easily embedded
in this space. The metric of the system concerned is given by
eq. (247) in \cite{Smith}, from which we conclude that the absence of
symmetry enhancement carries over into the strong coupling limit. 
Nonetheless, it is important to note that the
chiral fermion at the intersection gains mass and is
displaced away from the intersection even in the T-dual picture. 
This is exploited in
\cite{Harvey1, Harvey2} for chiral symmetry breaking.

Finally, we note that the symmetry enhancement disappears as soon as
the D5/D5 system is compactified on a circle since the two sets of D5
branes are no longer indistinguishable. This is T-dual to the
D6/D4 system on a circle in the D6 world-volume. The analysis here
suggests that compactification \emph{alone} destroys the symmetry enhancement.

\section*{Acknowledgement}
I would like to thank Prof. M. Green. for his guidance and
encouragements. I would also
like to thank Dr. D. Tong for his insightful comments.

\end{document}